\documentclass[aps,onecolumn,showpacs,showkeys,nofootinbib]{revtex4}
\usepackage{epsfig}
\usepackage{amsmath}
\usepackage{amsfonts}
\usepackage{amssymb}
\usepackage{graphicx}
\usepackage{colordvi}
\begin{document}
\begin{flushleft}
KCL-PH-TH/2014-47
\end{flushleft}

\title{Probing  Models of Extended Gravity using Gravity Probe B and LARES experiments}

\author{S. Capozziello$^{1,2,3}$\footnote{e - mail address: capozziello@na.infn.it}, G. Lambiase$^{4,5}$\footnote{e - mail address: lambiase@sa.infn.it}, M. Sakellariadou$^6$\footnote{e - mail address: mairi.sakellariadou@kcl.ac.uk},  An. Stabile$^{4,5}$\footnote{e - mail address: anstabile@gmail.com},
Ar. Stabile$^7$\footnote{e - mail address: arturo.stabile@gmail.com},}
\affiliation{$^1$Dipartimento di Fisica, Universit\`{a} di Napoli ``Federico II'', Complesso Universitario di Monte Sant'Angelo, Edificio G, Via Cinthia, I-80126, Napoli, Italy}
\affiliation{$^2$Istituto Nazionale di Fisica Nucleare (INFN) Sezione di Napoli, Complesso Universitario  di Monte Sant'Angelo, Edificio G, Via Cinthia, I-80126, Napoli, Italy}
\affiliation{$^3$ Gran Sasso Science Institute (INFN),  Viale F. Crispi, 7, I-67100, L'Aquila, Italy}
\affiliation{$^4$Dipartimento di Fisica ``E.R. Caianiello'', Universit\`{a} degli Studi di Salerno, via G. Paolo II, Stecca 9, I - 84084 Fisciano, Italy}
\affiliation{$^5$Istituto Nazionale di Fisica Nucleare (INFN) Sezione di Napoli, Gruppo collegato di Salerno}
\affiliation{$^6$Department of Physics, King's College London, University of London, Strand WC2R 2LS, London, United Kingdom}
\affiliation{$^7$Dipartimento di Ingegneria, Universit\`{a} del Sannio, Palazzo Dell'Aquila Bosco Lucarelli, Corso Garibaldi, 107 - 82100, Benevento, Italy}

\begin{abstract}
We consider models of Extended Gravity and in particular, generic models containing scalar-tensor and  higher-order curvature  terms, as well as a model derived from  noncommutative spectral geometry.
Studying, in the weak-field approximation, the geodesic and  Lense-Thirring processions, we impose constraints on the free parameters of such models by using the recent experimental results of the Gravity Probe B and LARES satellites.  
\end{abstract}
\date{\today}
\pacs{04.50.Kd; 04.25.Nx; 04.80.Cc }
\keywords{Modified theories of gravity;  post-Newtonian approximation; experimental tests of  gravitational theories.}
\maketitle

\section{Introduction}

Extended Gravity may offer an alternative approach  to explain cosmic acceleration  and large scale structure without considering  dark energy and dark matter. In this framework, while the well-established results of General Relativity (GR) are retained at local scales, deviations at  ultraviolet and infrared scales are considered~\cite{PRnostro}. In such models of Extended Gravity, which may result from some effective theory aiming at providing a full quantum gravity formulation, the gravitational interaction may contain further contributions, with respect to GR, at galactic, extra-galactic and  cosmological scales where, otherwise, large amounts of unknown dark components are required.

In the simplest version of Extended Gravity, the Ricci curvature scalar $R$, linear in the Hilbert-Einstein action, could be replaced by a generic function $f(R)$ whose true form could be
``reconstructed'' by the data. Indeed, in the absence of a full theory of Quantum Gravity, one may adopt the approach that observational data could contribute to define and constrain the ``true'' theory of gravity~\cite{PRnostro,reviewodi,reviewodi1,reviewmauro,reviewvalerio,libro,libro1}.

In the weak-field approximation,  any relativistic theory of gravitation  yields, in general, corrections to the gravitational potentials ({\em e.g.}, Ref.~\cite{Qua91}) which, at the  post-Newtonian  level and in the  Parametrized Post-Newtonian formalism, could constitute the test-bed for  these theories~\cite{Will93}.
In Extended Gravity there are further gravitational degrees of freedom (related to higher order terms, nonminimal couplings and scalar fields in the field equations), and moreover gravitational interaction is {\it not} invariant at any scale. Hence, besides the Schwarzschild radius, other characteristic gravitational scales could come out from dynamics. Such scales, in the weak field approximation, should be responsible for characteristic lengths of astrophysical structures that should result {\it confined} in this way~\cite{annalen}.
Considering gravity  at  local and microscopic level,  the possible violation of Equivalence Principle could open the door to test such additional degrees of freedom~\cite{stequest}.

In what follows, we investigate in Sec.\ \ref{STFOG_par}A the weak-field limit of generic scalar-tensor-higher-order models, in view of constraining their parameters by satellite data like Gravity Probe B and LARES.
In addition, we consider in Sec.\ \ref{STFOG_par}B a scalar-tensor-higher-order model derived from Noncommutative Spectral Geometry. The analysis is performed, in Sec.\ \ref{WFSTFOG_par}, in the Newtonian limit, and the solutions are found for a point-like source in Sec.\ \ref{PLSTFOG_par}), and for a rotating ball-like source in Sec.\ \ref{BLSTFOG_par}. In Sec.\ \ref{CRCSTFOG_par}, we review the aspects on circular rotatation curves and discuss the effects of the  parameters of the considered models. In the Sec.\ \ref{RSPSTFOG_par},  we analyze all orbital parameters for the case of a rotating source. The comparison with the experimental data is performed in Sec. \ref{EXSTFOG_par} and our conclusions are drawn in Sec.\ \ref{conclusions}.

\section{Extended Gravity}
\label{STFOG_par}
We will discuss the general case of scalar-tensor-higher-order gravity where the standard Hilbert-Einstein action is replaced by a more general action containing a scalar field and curvature invariants,  like the Ricci scalar $R$ and the Ricci tensor $R_{\alpha\beta}$. We note that the Riemann tensor can be discarded since the Gauss-Bonnet invariant fixes it in the action (for details see Ref.~\cite{cqg}). We derive the field equations and, in particular, discuss the case of Noncommutative Geometry in order to show that such an approach is well-founded at the relevant scales.

\subsection{The general case: scalar-tensor-higher-order gravity}
Consider the action
\begin{eqnarray}\label{FOGaction}
\mathcal{S}\,=\,\int d^{4}x\sqrt{-g}\biggl[f(R,R_{\alpha\beta}R^{\alpha\beta},\phi)+\omega(\phi)\phi_{;\alpha}\phi^{;\alpha}+\mathcal{X}\mathcal{L}_m\biggr]~,
\end{eqnarray}
where $f$ is an unspecified function of the Ricci scalar $R$, the curvature  invariant $R_{\alpha\beta}R^{\alpha\beta}\,\doteq\,Y$ where $R_{\alpha\beta}$ is the Ricci scalar, and a scalar field $\phi$. Here $\mathcal{L}_m$ is the minimally coupled ordinary matter Lagrangian density, $\omega$ is a generic function of the scalar field, $g$ is the determinant of metric tensor $g_{\mu\nu}$ and\footnote{Here we use the convention $c\,=\,1$.} $\mathcal{X}\,=\,8\pi G$. In the metric approach, namely when  the gravitational field is fully described by the metric tensor $g_{\mu\nu}$ only\footnote{It is worth noticing that in metric-affine theories, the gravitational field is completely assigned by the metric tensor $g_{\mu\nu}$, while the affinity connections $\Gamma^{\alpha}_{\mu\nu}$ are considered as independent fields~\cite{PRnostro}.}, the field equations are obtained by varying the action (\ref{FOGaction}) with respect to $g_{\mu\nu}$, leading to
\begin{eqnarray}\label{fieldequationFOG}
&&f_RR_{\mu\nu}-\frac{f+\omega(\phi)\phi_{;\alpha}\phi^{;\alpha}}{2}g_{\mu\nu}-f_{R;\mu\nu}+g_{\mu\nu}\Box
f_R+2f_Y{R_\mu}^\alpha
R_{\alpha\nu}
\nonumber\\\\\nonumber&&-2[f_Y{R^\alpha}_{(\mu}]_{;\nu)\alpha}+\Box[f_YR_{\mu\nu}]+[f_YR_{\alpha\beta}]^{;\alpha\beta}g_{\mu\nu}+\omega(\phi)\phi_{;\mu}\phi_{;\nu}\,=\,
\mathcal{X}\,T_{\mu\nu}~,
\end{eqnarray}
where $T_{\mu\nu}\,=\,-\frac{1}{\sqrt{-g}}\frac{\delta(\sqrt{-g}\mathcal{L}_m)}{\delta
g^{\mu\nu}}$ is the the energy-momentum tensor of matter, $f_R\,=\,\frac{df}{dR}$, $f_Y\,=\,\frac{df}{dY}$ and $\Box={{}_{;\sigma}}^{;\sigma}$ is the D'Alembert operator. We use for the Ricci tensor the convention 
$R_{\mu\nu}={R^\sigma}_{\mu\sigma\nu}$, whilst for the Riemann
tensor we define ${R^\alpha}_{\beta\mu\nu}=\Gamma^\alpha_{\beta\nu,\mu}+\cdots$. The
affinity connections are the usual Christoffel symbols of the metric, namely
$\Gamma^\mu_{\alpha\beta}=\frac{1}{2}g^{\mu\sigma}(g_{\alpha\sigma,\beta}+g_{\beta\sigma,\alpha}
-g_{\alpha\beta,\sigma})$, and we adopt the signature is $(+,-,-,-)$. The trace of the field equation
Eq.~(\ref{fieldequationFOG}) above, reads
\begin{eqnarray}\label{tracefieldequationFOG}
f_RR+2f_YR_{\alpha\beta}R^{\alpha\beta}-2f+\Box[3
f_R+f_YR]+2[f_YR^{\alpha\beta}]_{;\alpha\beta}-\omega(\phi)\phi_{;\alpha}\phi^{;\alpha}\,=\,\mathcal{X}\,T~,\end{eqnarray}
where $T\,=\,T^{\sigma}_{\,\,\,\,\,\sigma}$ is the trace of
energy-momentum tensor. 

By varying the action (\ref{FOGaction}) with respect to the scalar field $\phi$,  we obtain the Klein-Gordon field equation
\begin{eqnarray}\label{FE_SF}
2\omega(\phi)\Box\phi+\omega_\phi(\phi)\phi_{;\alpha}\phi^{;\alpha}-f_\phi\,=\,0~,
\end{eqnarray}
where $\omega_\phi(\phi)\,=\,\frac{d\omega(\phi)}{d\phi}$ and $f_\phi\,=\,\frac{df}{d\phi}$.

In the following sub-section we will consider a particular model derived by a fundamental theory, namely by noncommutative spectral geometry~\cite{ncg-book1,ncg-book2}.

\subsection{The case of Noncommutative Spectral Geometry}
\label{NCG}
\label{NCSG}
Running backwards in time the evolution of our universe, we approach extremely high energy scales and huge densities within tiny spaces. At such extreme conditions, GR can no longer describe satisfactorily the underlined physics, and a full Quantum Gravity  Theory has to be invoked. Different Quantum Gravity approaches have been  worked out in the literature; they should all lead to GR, considered as an effective theory, as one  reaches energy scales much below the Planck scale.

Even though Quantum Gravity may imply that at Planck energy scales spacetime is a widly noncommutative manifold, one may safely assume that at scales a few orders of magnitude below the Planck scale, the spacetime is only mildy noncommutative. At such intermediate scales, the algebra of coordinates can be considered as an almost-commutative algebra of matrix valued functions, which if appropriately chosen, can lead to the Standard Model  of particle physics. The application of the spectral action principle~\cite{Chamseddine:1996zu} to this almost-commutative manifold led to the NonCommutative Spectral Geometry (NCSG)~\cite{Sakellariadou:2010nr,Sakellariadou:2012jz,vandenDungen:2012ky}, a framework that offers  a purely geometric explanation of the Standard Model of particles coupled to gravity~\cite{ccm,cchiggs}.

 For almost-commutative manifolds, the geometry is described by the tensor product ${\cal M}\times {\cal F}$ of
a four-dimensional compact Riemannian manifold ${\cal M}$ and
a discrete noncommutative space ${\cal F}$, with ${\cal M}$ describing the geometry of spacetime and ${\cal F}$ the internal space of the particle physics model.
The noncommutative nature of ${\cal F}$ is encoded in the spectral
triple $\left({\cal A}_{\cal F},{\cal H}_{\cal F}, D_{\cal F}\right)$.
The algebra ${\cal A}_{\cal F}=C^\infty({\cal M})$ of smooth functions
on ${\cal M}$, playing the r\^ole
of the algebra of coordinates, is an involution of operators on the finite-dimensional
Hilbert space ${\cal H_F}$ of Euclidean fermions.  The operator $D_{\cal F}$ is the Dirac
operator ${\partial\hspace{-5pt}\slash}_{\cal
  M}=\sqrt{-1}\gamma^\mu\nabla_\mu^s$ on the spin manifold ${\cal M}$;
it corresponds to the inverse of the Euclidean propagator of fermions
and is given by the Yukawa coupling matrix and the Kobayashi-Maskawa mixing parameters.

The  algebra ${\cal A}_{\cal F}$ has to be chosen so that it can lead to the Standard Model of particle physics, while it must also fulfill noncommutative geometry requirements. It was hence chosen to be~\cite{Chamseddine:2007ia,Sakellariadou:2011wv,Gargiulo:2013bla}
\begin{eqnarray}
\nonumber
{\cal A}_{\cal F}=M_{a}(\mathbb{H})\oplus M_{k}(\mathbb{C})~,
\end{eqnarray}
with $k=2a$; $\mathbb{H}$ is the algebra of quaternions, which encodes
the noncommutativity of the manifold.  The first possible value for
$k$ is 2, corresponding to a Hilbert space of four fermions; it is ruled out from the existence of quarks.
The minimum possible value for $k$ is 4 leading to the correct number of $k^2=16$
fermions in each of the three generations. Higher values of $k$ can lead to particle physics models
beyond the Standard Model~\cite{Devastato:2013oqa,Chamseddine:2013rta}.
The spectral geometry in the product ${\cal M}\times {\cal F}$  is
given by the product rules:
\begin{eqnarray}
{\cal A} &=& C^\infty({\cal M})\oplus{\cal A_F}\ , \nonumber\\
  {\cal H}&=&L^2({\cal M},S)\oplus{\cal H_F}\ , \nonumber\\
{\cal D}&=&{\cal D_M}\oplus1+\gamma_5\oplus{\cal D_F}~,
\end{eqnarray}
where
$L^2({\cal M}, S)$ is the Hilbert space of $L^2$ spinors and ${\cal
D_M}$ is the Dirac operator of the Levi-Civita spin connection on
${\cal M}$.
Applying the spectral action principle to
the product geometry ${\cal M}\times {\cal F}$ leads to the NCSG action
\begin{eqnarray}
\nonumber
{\rm Tr}(f(D_{\cal A}/\Lambda))+(1/2)\langle J\psi,D\psi\rangle~,
\end{eqnarray}
splitted into the bare bosonic action and the fermionic one.
Note that $D_{\cal A}=D +{\cal A}+\epsilon'J{\cal A}J^{-1}$ are
uni-modular inner fluctuations, $f$ is a cutoff function and $\Lambda$
fixes the energy scale, $J$ is the real structure on the spectral
triple and $\psi$ is a spinor in the Hilbert space ${\cal H}$ of the
quarks and leptons.
In what follows we  concentrate on the bosonic part of the action, seen
 as the bare action at the mass scale $\Lambda$ which
includes the eigenvalues of the Dirac
operator that are smaller than the cutoff scale $\Lambda$, considered as the grand unification scale.
 Using heat
kernel methods, the trace ${\rm Tr}(f({\cal D}_A/\Lambda)$ can be
written in terms of the geometrical Seeley-de Witt coefficients $a_n$ as~\cite{Chamseddine:2005zk,Chamseddine:2008zj}
\begin{eqnarray}
\label{asymp-exp} {\rm Tr}(f({\cal D}_{\cal A}/\Lambda))\sim
2\Lambda^4f_4a_0+2\Lambda^2f_2a_2+f_0a_4+\cdots
+\Lambda^{-2k}f_{-2k}a_{4+2k}+\cdots~,
\end{eqnarray}
with $f_k$ the momenta of the smooth even test (cutoff) function which decays fast at infinity:
\begin{eqnarray}
\label{eq:moments0}
f_0 &\equiv& f(0)~, \nonumber\\
f_k &\equiv&\int_0^\infty f(u) u^{k-1}{\rm
  d}u\ \ ,\ \ \mbox{for}\ \ k>0 ~,\nonumber\\ \mbox
    f_{-2k}&=&(-1)^k\frac{k!}{(2k)!} f^{(2k)}(0)~.  \nonumber
\end{eqnarray}
Since the Taylor expansion of the $f$ function vanishes at zero, the
asymptotic expansion of the spectral action reduces to
\begin{eqnarray}
\label{asympt} {\rm Tr}(f({\cal D}_{\cal A}/\Lambda))\sim
2\Lambda^4f_4a_0+2\Lambda^2f_2a_2+f_0a_4~.
\end{eqnarray}
Hence, the cutoff function $f$ plays a r\^ole only through its momenta
$f_0, f_2, f_4$, three real parameters, related to the coupling
constants at unification, the gravitational constant, and the
cosmological constant, respectively.

The NCSG model lives by construction at the grand unification scale, hence
providing a framework to study early universe
cosmology~\cite{Nelson:2008uy,Nelson:2009wr,Marcolli:2009in,Buck:2010sv}.  The
gravitational part of the asymptotic expression for the bosonic sector of
the NCSG action\footnote{Note that the obtained action does not suffer fot negative energy massive graviton modes~\cite{Donoghue:1994dn}.}, including the coupling between the Higgs field $\phi$
and the Ricci curvature scalar $R$, in Lorentzian signature,
obtained through a Wick rotation in imaginary time, reads~\cite{ccm}
\begin{equation}
\label{eq:0}
\mathcal{S}_{\rm grav}^{\rm L}= \int  {\rm d}^4 x\sqrt{-g}\left[\frac{R}{2\kappa_0^2} + \alpha_0
C_{\alpha\beta\gamma\delta}C^{\alpha\beta\gamma\delta} + \tau_0 R^\star
R^\star - \xi_0 R|{\bf H}|^2 \right]\,;
\end{equation}
${\bf H}=(\sqrt{af_0}/\pi)\phi$, with $a$ a parameter related to
fermion and lepton masses and lepton mixing. At unification scale (set
up by $\Lambda$), $\alpha_0=-3f_0/(10\pi^2)$, $\xi_0=\frac{1}{12}$.

The square of the Weyl tensor can be expressed in terms of $R^2$ and $R_{\alpha\beta}R^{\alpha\beta}$ as
\begin{eqnarray}
\nonumber
C_{\alpha\beta\gamma\delta}C^{\alpha\beta\gamma\delta}\,=\,2R_{\alpha\beta}R^{\alpha\beta}-\frac{2}{3}R^2.
\end{eqnarray}
The above action (\ref{eq:0}) is clearly a particular case of the action (\ref{FOGaction}) describing a general model of an Extended Theory of Gravity. As we will show in the following, it may lead to effects observable at local scales (in particular at Solar System scales), hence it may be tested against current
gravitational data.


\section{The Weak-Field Limit }
\label{WFSTFOG_par}

We will  study, in the weak-field approximation, models of Extended Gravity at Solar System scales. 
In order to perform the weak-field limit, we have to perturb Eqs.~(\ref{fieldequationFOG}), (\ref{tracefieldequationFOG}) and (\ref{FE_SF}) in a  Minkowski background $\eta_{\mu\nu}$~\cite{noi-newt,mio1}. We set
\begin{eqnarray}\label{PM_me}
&&g_{\mu\nu}\,\sim\,\begin{pmatrix}
1+g^{(2)}_{tt}(t,\mathbf{x})+g^{(4)}_{tt}(t,\mathbf{x})+\dots & g^{(3)}_{ti}(t,\mathbf{x})+\dots \\
g^{(3)}_{ti}(t,\mathbf{x})+\dots & -\delta_{ij}+g^{(2)}_{ij}(t,\mathbf{x})+\dots\end{pmatrix}\,=\,
\begin{pmatrix}
1+2\Phi+2\Xi & 2A_i \\
2A_i & -\delta_{ij}+2\Psi\delta_{ij}\end{pmatrix}~,
\nonumber\\\\
&&\phi\,\sim\,\phi^{(0)}+\phi^{(2)}+\dots\,=\,\phi^{(0)}+\varphi~,\nonumber
\end{eqnarray}
where $\Phi$, $\Psi$, $\varphi$ are proportional to the power $(v/c)^2$ (Newtonian limit) while $A_i$ is proportional to $(v/c)^3$ and $\Xi$ to $(v/c)^4$ (post-Newtonian limit).
The function $f$,  up to  the $(v/c)^3$ order,  can be developed as

\begin{eqnarray}
f(R,R_{\alpha\beta}R^{\alpha\beta},\phi)\,=\,&&f_{R}(0,0,\phi^{(0)})\,R+\frac{f_{RR}(0,0,\phi^{(0)})}{2}\,R^2+\frac{f_{\phi\phi}(0,0,\phi^{(0)})}{2}(\phi-\phi^{(0)})^2\nonumber\\\\\nonumber&&+f_{R\phi}(0,0,\phi^{(0)})R\,\phi+f_Y(0,0,\phi^{(0)})R_{\alpha\beta}R^{\alpha\beta}~,
\end{eqnarray}
while all other possible contributions in $f$ are negligible \cite{mio1,mio2,FOGST}. The field equations (\ref{fieldequationFOG}), (\ref{tracefieldequationFOG}) and (\ref{FE_SF}) hence read
\begin{eqnarray}\label{PMfieldequationFOG}
\begin{array}{ll}
f_{R}(0,0,\phi^{(0)})\,\biggl[R_{tt}-\frac{R}{2}\biggr]-f_Y(0,0,\phi^{(0)})\triangle R_{tt}-[f_{RR}(0,0,\phi^{(0)})+\frac{f_Y(0,0,\phi^{(0)})}{2}]\triangle
R-f_{R\phi}(0,0,\phi^{(0)})\triangle\varphi\,=\,\mathcal{X}\,T_{tt}~,\\\\
f_{R}(0,0,\phi^{(0)})\,\biggl[R_{ij}+\frac{R}{2}\delta_{ij}\biggr]-f_Y(0,0,\phi^{(0)})\triangle R_{ij}+[f_{RR}(0,0,\phi^{(0)})+\frac{f_Y(0,0,\phi^{(0)})}{2}]\delta_{ij}\triangle
R-f_{RR}(0,0,\phi^{(0)})R_{,ij}\\\\\qquad\qquad\qquad-2f_Y(0,0,\phi^{(0)})R^\alpha_{\,\,\,\,(i,j)\alpha}-f_{R\phi}(0,0,\phi^{(0)})(\partial^2_{ij}-\delta_{ij}\triangle)\varphi\,=\,\mathcal{X}\,T_{ij}~,
\\\\
f_{R}(0,0,\phi^{(0)})\,R_{ti}-f_Y(0,0,\phi^{(0)})\triangle R_{ti}-f_{RR}(0,0,\phi^{(0)})R_{,ti}-2f_Y(0,0,\phi^{(0)})R^\alpha_{\,\,\,\,(t,i)\alpha}-f_{R\phi}(0,0,\phi^{(0)})\,\varphi_{,ti}\,=\,\mathcal{X}\,T_{ti}~,
\\\\
f_{R}(0,0,\phi^{(0)})\,R+[3f_{RR}(0,0,\phi^{(0)})+2f_Y(0,0,\phi^{(0)})]\triangle
R+3f_{R\phi}(0,0,\phi^{(0)})\triangle\varphi\,=\,-\mathcal{X}\,T~,\\\\
2\omega(\phi^{(0)})\triangle\varphi+f_{\phi\phi}(0,0,\phi^{(0)})\varphi+f_{R\phi}(0,0,\phi^{(0)})R\,=\,0~,
\end{array}
\end{eqnarray}
where $\triangle$ is the Laplace operator in the flat space. The geometric quantities $R_{\mu\nu}$ and  $R$ are evaluated at the first order with respect to  the metric potentials $\Phi$, $\Psi$ and $A_i$. By introducing the quantities\footnote{In the Newtonian and post-Newtonian limits,  we can consider as Lagrangian in the action (\ref{FOGaction}),  the quantity $f(X,Y)\,=\,a\,R+b\,R^2+c\,R_{\alpha\beta}R^{\alpha\beta}$~\cite{mio2}. Then the masses (\ref{mass_definition}) become ${m_R}^2\,=\,-\frac{a}{2(3b+c)}$, ${m_Y}^2\,=\,\frac{a}{c}$. For a correct interpretation of these quantities as real masses, we have to impose $a\,>\,0$, $b\,<\,0$ and $0\,<\,c\,<\,-3b$.}

\begin{eqnarray}\label{mass_definition}
\begin{array}{ll}
{m_R}^2\,\doteq\,-\frac{f_{R}(0,0,\phi^{(0)})}{3f_{RR}(0,0,\phi^{(0)})+2f_Y(0,0,\phi^{(0)})}~,\\\\
{m_Y}^2\,\doteq\,\frac{f_{R}(0,0,\phi^{(0)})}{f_Y(0,0,\phi^{(0)})}~,\\\\
{m_\phi}^2\,\doteq\,-\frac{f_{\phi\phi}(0,0,\phi^{(0)})}{2\omega(\phi^{(0)})}~,
\end{array}
\end{eqnarray}
and setting $f_R(0,0,\phi^{(0)})\,=\,1$, $\omega(\phi^{(0)})\,=\,1/2$ for simplicity\footnote{We can define a new gravitational constant: $\mathcal{X}\,\rightarrow\,\mathcal{X}\,f_R(0,0,\phi^{(0)})$ and $f_{R\phi}(0,0,\phi^{0})\,\rightarrow\,f_{R\phi}(0,0,\phi^{0})\,f_R(0,0,\phi^{(0)})$.}, we get the complete set of differential equations

\begin{eqnarray}
\begin{array}{ll}
(\triangle-{m_Y}^2)R_{tt}+\biggl[\frac{{m_Y}^2}{2}-\frac{{m_R}^2+2{m_Y}^2}{6{m_R}^2}\triangle\biggr]
R+{m_Y}^2\,f_{R\phi}(0,0,\phi^{(0)})\,\triangle\varphi\,=\,-{m_Y}^2\mathcal{X}\,T_{tt}~,\\ \\
(\triangle-{m_Y}^2)R_{ij}+\biggl[\frac{{m_R}^2-{m_Y}^2}{3{m_R}^2}\,
\partial^2_{ij}-\delta_{ij}\biggl(\frac{{m_Y}^2}{2}-\frac{{m_R}^2+2{m_Y}^2}{6{m_R}^2}\triangle\biggr)\biggr]R \\ \\
\qquad \qquad \qquad \qquad +{m_Y}^2\,f_{R\phi}(0,0,\phi^{(0)})\,(\partial^2_{ij}-\delta_{ij}\triangle)\varphi\,=\,-{m_Y}^2\,\mathcal{X}\,T_{ij}~,
\\ \\
(\triangle-{m_Y}^2)R_{ti}+\frac{{m_R}^2-{m_Y}^2}{3{m_R}^2}R_{,ti}+{m_Y}^2\,f_{R\phi}(0,0,\phi^{(0)})\,\varphi_{,ti}\,=\,-{m_Y}^2\mathcal{X}\,T_{ti}~,
\label{PMfieldequationFOG2}
\\ \\
(\triangle-{m_R}^2)R-3{m_R}^2\,f_{R\phi}(0,0,\phi^{(0)})\,\triangle\varphi\,=\,{m_R}^2\,\mathcal{X}\,T~,\\\\
(\triangle-{m_\phi}^2)\varphi+f_{R\phi}(0,0,\phi^{(0)})\,R\,=\,0~.
\end{array}
\end{eqnarray}
The components of the Ricci tensor in Eq.~(\ref{PMfieldequationFOG2})  in the weak-field limit read 
\begin{eqnarray}\label{ricci_tensor_components}
\begin{array}{ll}
R_{tt}\,=\,\frac{1}{2}\triangle\,g^{(2)}_{tt}\,=\,\triangle\Phi~,\\\\
R_{ij}\,=\,\frac{1}{2}g^{(2)}_{ij,mm}-\frac{1}{2}g^{(2)}_{im,mj}-\frac{1}{2}g^{(2)}_{jm,mi}-\frac{1}{2}g^{(2)}_
{tt,ij}+\frac{1}{2}g^{(2)}_{mm,ij}\,=\,\triangle\Psi\,\delta_{ij}+(\Psi-\Phi)_{,ij}~,\\\\
R_{ti}=\frac{1}{2}g^{(3)}_{ti,mm}-\frac{1}{2}g^{(2)}_{im,mt}-\frac{1}{2}g^{(3)}_{mt,mi}+\frac{1}{2}g^{(2)}_
{mm,ti}\,=\,\triangle A_i+\Psi_{,ti}~.
\end{array}
\end{eqnarray}
The energy momentum tensor $T_{\mu\nu}$ can be also expanded. For a perfect fluid, when the pressure is negligible with respect to the mass density $\rho$, it reads $T_{\mu\nu}\,=\,\rho\,u_\mu u_\nu$ with $u_\sigma u^\sigma\,=\,1$. However, the development starts form the zeroth order\footnote{This formalism descends from the theoretical setting of Newtonian mechanics which requires the appropriate scheme of approximation when obtained from a more general relativistic theory. This scheme coincides with a gravity theory analyzed at the first order of perturbation in a curved spacetime metric.}, hence $T_{tt}\,=\,T^{(0)}_{tt}\,=\,\rho$, $T_{ij}\,=\,T^{(0)}_{ij}\,=\,0$ and $T_{ti}\,=\,T^{(1)}_{ti}\,=\,\rho\,v_i$, where $\rho$ is the density mass and $v^i$ is the velocity of the source. Thus, $T_{\mu\nu}$ is independent of metric potentials and satisfies the ordinary conservation condition $T^{\mu\nu}_{\,\,\,\,\,\,\,\,\,,\mu}\,=\,0$. Equations~(\ref{PMfieldequationFOG2}) thus read
\begin{eqnarray}\label{PMfieldequationFOG3}
\begin{array}{ll}
(\triangle-{m_Y}^2)\triangle\Phi+\biggl[\frac{{m_Y}^2}{2}-\frac{{m_R}^2+2{m_Y}^2}{6{m_R}^2}\triangle\biggr]
R+{m_Y}^2\,f_{R\phi}(0,0,\phi^{(0)})\,\triangle\varphi\,=\,-{m_Y}^2\mathcal{X}\,\rho~,\\\\
\biggl\{(\triangle-{m_Y}^2)\triangle\Psi-\biggl[\frac{{m_Y}^2}{2}-\frac{{m_R}^2+2{m_Y}^2}{6{m_R}^2}\triangle\biggr]R-{m_Y}^2\,f_{R\phi}(0,0,\phi^{(0)})\,\triangle\varphi\biggr\}\delta_{ij}\\\\\qquad\qquad\qquad\qquad\qquad\qquad+\biggl\{(\triangle-{m_Y}^2)(\Psi-\Phi)+\frac{{m_R}^2-{m_Y}^2}{3{m_R}^2}\,R+{m_Y}^2\,f_{R\phi}(0,0,\phi^{(0)})\,\varphi\biggr\}_{,ij}\,=\,0~,
\\\\
\biggl\{(\triangle-{m_Y}^2)\triangle A_i+{m_Y}^2\mathcal{X}\,\rho\,v_i\biggr\}+\biggl\{(\triangle-{m_Y}^2)\Psi+\frac{{m_R}^2-{m_Y}^2}{3{m_R}^2}R+{m_Y}^2\,f_{R\phi}(0,0,\phi^{(0)})\,\varphi\biggr\}_{,ti}\,=\,0~,
\\\\
(\triangle-{m_R}^2)R-3{m_R}^2\,f_{R\phi}(0,0,\phi^{(0)})\,\triangle\varphi\,=\,{m_R}^2\,\mathcal{X}\,\rho~,\\\\
(\triangle-{m_\phi}^2)\varphi+f_{R\phi}(0,0,\phi^{(0)})\,R\,=\,0~.
\end{array}
\end{eqnarray}
In the following we will consider the Newtonian and Post-Newtonian limits.

\subsection{The Newtonian limit:  solutions of the fields $\Phi$, $\varphi$ and $R$}\label{PLSTFOG_par}
Equations~(\ref{PMfieldequationFOG3}a) and (\ref{PMfieldequationFOG3}b)  are coupled system and, for a point-like source $\rho(\mathbf{x})\,=\,M\,\delta(\mathbf{x})$, admit the solutions:

\begin{eqnarray}
\label{ST_FOG_FE_NL_sol_sys}
\begin{array}{ll}
\varphi(\textbf{x})\,=\,\sqrt{\frac{\xi}{3}}\,\frac{r_g}{|\textbf{x}|}\frac{e^{-m_R\tilde{k}_R\,|\textbf{x}|}-e^{-m_R\tilde{k}_\phi\,|\textbf{x}|}}{{\tilde{k}_R}^2-{\tilde{k}_\phi}^2}~,\\\\
R(\textbf{x})\,=\,-{m_R}^2\frac{r_{\rm g}}{|\textbf{x}|}\frac{({\tilde{k}_R}^2-\eta^2)\,e^{-m_R\tilde{k}_R\,|\textbf{x}|}-({\tilde{k}_\phi}^2-\eta^2)\,e^{-m_R\tilde{k}_\phi\,|\textbf{x}|}}{{\tilde{k}_R}^2-{\tilde{k}_\phi}^2}~,
\end{array}
\end{eqnarray}
where $r_{\rm g}$ is the Schwarzschild radius, ${\tilde{k}_{R,\phi}}^2\,=\,\frac{1-\xi+\eta^2\pm\sqrt{(1-\xi+\eta^2)^2-4\eta^2}}{2}$, $\xi\,=\,3{f_{R\phi}(0,0,\phi^{(0)})}^2$ and $\eta\,=\,\frac{m_\phi}{m_R}$ \cite{FOGST}\footnote{The parameter $\xi$ is defined generally as $\frac{3{f_{R\phi}(0,0,\phi^{(0)})}^2}{2\,f_R(0,0,\phi^{(0)})\,\omega(\phi^{(0)})}$.}. Moreover $\xi$ and $\eta$ satisfy the condition $(\eta-1)^2-\xi\,>\,0$. The formal solution of the gravitational potential $\Phi$, derived from Eq.~(\ref{PMfieldequationFOG3}a), reads 

\begin{eqnarray}\label{ST_FOG_FE_NL_sol}
\Phi(\mathbf{x})\,=\,\frac{-1}{16\pi^2}\int
\frac{d^3\mathbf{x}'d^3\mathbf{x}''}{|\mathbf{x}-\mathbf{x}'|}\frac{e^{-m_Y|\mathbf{x}'-\mathbf{x}''|}}{|\mathbf{x}'-\mathbf{x}''|}\biggl[
\frac{4{m_Y}^2-{m_R}^2}{6}\mathcal{X}\,\rho(\mathbf{x}'')+\frac{{m_Y}^2-{m_R}^2(1-\xi)}{6}R(\mathbf{x}'')-\frac{{m_R}^4\eta^2}{2\sqrt{3}}\,\xi^{1/2}\,\varphi(\mathbf{x}'')\biggr]~,
\nonumber\end{eqnarray}
which for a point-like source is

\begin{eqnarray}\label{ST_FOG_FE_NL_sol_point}
\Phi(\mathbf{x})\,=\,-\frac{GM}{|\mathbf{x}|}\biggl[1+g(\xi,\eta)\,e^{-m_R\tilde{k}_R|\mathbf{x}|}+[\frac{1}{3}-g(\xi,\eta)]\,e^{-m_R\tilde{k}_\phi|\mathbf{x}|}-\frac{4}{3}\,e^{-m_Y|\mathbf{x}|}\biggr]~,
\end{eqnarray}
where 
 \[
 g(\xi,\eta)\,=\,\frac{1-\eta^2+\xi+\sqrt{\eta^4+(\xi-1)^2-2\eta^2(\xi+1)}}{6\sqrt{\eta^4+(\xi-1)^2-2\eta^2(\xi+1)}}\,~.
  \]
Note that for $f_Y\,\rightarrow\,0$ \emph{i.e.} $m_Y\,\rightarrow\,\infty$, we obtain the same outcome for the gravitational potential as in Ref.~\cite{FOGST} for a $f(R,\phi)$-theory. The absence of the coupling term between the curvature invariant $Y$ and the scalar field $\phi$, as well as the linearity of the field equations (\ref{PMfieldequationFOG3}) guarantee that the solution (\ref{ST_FOG_FE_NL_sol_point}) is a linear combination of solutions obtained within an $f(R,\phi)$-theory and an $R+Y/{m_Y}^2$-theory. 

\subsection{The Post-Newtonian limit:  solutions of the fields $\Psi$ and $A_i$}\label{BLSTFOG_par}
Equation~(\ref{PMfieldequationFOG3}b) can be formally solved as

\begin{eqnarray}\label{sol_psi_formally}
\Psi(\mathbf{x})\,=\,\Phi(\mathbf{x})+\frac{{m_R}^2-{m_Y}^2}{12\pi{m_R}^2}\int d^3\mathbf{x}'\frac{e^{-m_Y|\mathbf{x}-\mathbf{x}'|}}{|\mathbf{x}-\mathbf{x}'|}\,R(\mathbf{x}')+\frac{{m_Y}^2\xi^{1/2}}{4\sqrt{3}\pi}\int d^3\mathbf{x}'\frac{e^{-m_Y|\mathbf{x}-\mathbf{x}'|}}{|\mathbf{x}-\mathbf{x}'|}\,\varphi(\mathbf{x}')~,\nonumber
\end{eqnarray}
which for a point-like source reads
\begin{eqnarray}\label{ST_FOG_FE_NL_sol_point_psi}
\Psi(\mathbf{x})\,=\,-\frac{GM}{|\mathbf{x}|}\biggl[1-g(\xi,\eta)\,e^{-m_R\tilde{k}_R|\mathbf{x}|}-[1/3-g(\xi,\eta)]\,e^{-m_R\tilde{k}_\phi|\mathbf{x}|}-\frac{2}{3}\,e^{-m_Y|\mathbf{x}|}\biggr]~,
\end{eqnarray}
obtained by setting $\{\dots\}_{,ij}\,=\,0$ in Eq.~(\ref{PMfieldequationFOG3}b), while one also has $\{\dots\}\delta_{ij}\,=\,0$ leading to

\begin{eqnarray}\label{sol_psi_formally_2}
&\Psi(\mathbf{x})=-\frac{1}{16\pi^2}\int d^3\mathbf{x}'d^3\mathbf{x}''\frac{e^{-m_Y|\mathbf{x}'-\mathbf{x}''|}}{|\mathbf{x}-\mathbf{x}'||\mathbf{x}'-\mathbf{x}''|}\biggl[\frac{{m_R}^2+2{m_Y}^2}{6}\mathcal{X}\rho(\mathbf{x}'')-\frac{{m_Y}^2-{m_R}^2(1-\xi)}{6}R(\mathbf{x}'')+\frac{{m_R}^4\eta^2}{2\sqrt{3}}\xi^{1/2}\varphi(\mathbf{x}'')\biggr]&~,\nonumber\\
&&
\end{eqnarray}
which is however equivalent to  solution (\ref{sol_psi_formally}). The solutions (\ref{ST_FOG_FE_NL_sol_point}) and (\ref{ST_FOG_FE_NL_sol_point_psi}) generalize the outcomes of the theory $f(R,\,R_{\alpha\beta}R^{\alpha\beta})$ \cite{mio2}.
 
From Eq.~(\ref{PMfieldequationFOG3}c), we immediately obtain the solution for $A_i$, namely

\begin{eqnarray}\label{vect_pot}
A_i(\mathbf{x})\,=\,-\frac{{m_Y}^2\mathcal{X}}{16\pi^2}\int
d^3\mathbf{x}'d^3\mathbf{x}''\,\frac{e^{-m_Y|\mathbf{x}'-\mathbf{x}''|}}{|\mathbf{x}-\mathbf{x}'||\mathbf{x}'-\mathbf{x}''|}\,\rho(\mathbf{x}'')\,v''_i~.
\end{eqnarray}
In Fourier space,  solution (\ref{vect_pot}) presents the massless pole of  General Relativity, and the massive one\footnote{Note that Eq.~(\ref{PMfieldequationFOG3}c) in Fourier space becomes $|\mathbf{k}|^2(|\mathbf{k}|^2+{m_Y}^2)\tilde{A}_i\,=\,-{m_Y}^2\mathcal{X}\tilde{T}_{ti}$ and its solution reads $\tilde{A}_i\,=\,
-\mathcal{X}\tilde{T}_{ti}\Biggl[\frac{1}{|\mathbf{k}|^2}-\frac{1}{|\mathbf{k}|^2+m_Y^2}\Biggr]$.} is induced by the presence of the $R_{\alpha\beta}R^{\alpha\beta}$ term. Hence, the solution (\ref{vect_pot}) can be rewritten as the sum of  General Relativity contributions and massive modes. Since we do not consider  contributions inside  rotating bodies, we obtain

\begin{eqnarray}\label{vect_pot2}
A_i(\mathbf{x})\,=\,-\frac{\mathcal{X}}{4\pi}\int d^3\mathbf{x}'\frac{\rho(\mathbf{x}')\,v'_i}{|\mathbf{x}-\mathbf{x}'|}+\frac{\mathcal{X}}{4\pi}\int d^3\mathbf{x}'\frac{e^{-m_Y|\mathbf{x}-\mathbf{x}'|}}{|\mathbf{x}-\mathbf{x}'|}\,\rho(\mathbf{x}')\,v'_i~.
\end{eqnarray}
For a spherically symmetric system ($|\mathbf{x}|\,=\,r$) at rest and rotating with angular frequency $\mathbf{\Omega}(r)$, the energy momentum tensor $T_{ti}$ is 

\begin{eqnarray}
T_{ti}\,=\,\rho(\mathbf{x})\,v_i\,=\,T_{tt}(r)\,[\mathbf{\Omega}(r)\times\mathbf{x}]_i\,=\,\frac{3M}{4\pi\mathcal{R}^3}\Theta(\mathcal{R}-r)\,[\mathbf{\Omega}(r)\times\mathbf{x}]_i~,
\end{eqnarray}
where $\mathcal{R}$ is the radius of the body and $\Theta$ is the Heaviside function. Since only in General Relativity and  Scalar Tensor Theories the Gauss theorem is satisfied, here we have to consider the potentials $\Phi$, $\Psi$ generated by the ball source with radius $\mathcal{R}$, while they also depend on the shape of the source. In fact for any term $\propto\,\frac{e^{-mr}}{r}$, there is a geometric factor multiplying the Yukawa term, namely $F(m\,\mathcal{R})\,=\,3\frac{m\,\mathcal{R} \cosh m\,\mathcal{R}-\sinh m\,\mathcal{R}}{m^3\mathcal{R}^3}$. We thus get

\begin{eqnarray}\label{ST_FOG_FE_NL_sol_ball}
\begin{array}{ll}
\Phi_{\rm ball}(\mathbf{x})\,=\,-\frac{GM}{|\mathbf{x}|}\biggl[1+g(\xi,\eta)\,F(m_R\tilde{k}_R\mathcal{R})\,e^{-m_R\tilde{k}_R|\mathbf{x}|}+[\frac{1}{3}-g(\xi,\eta)]\,F(m_R\tilde{k}_\phi\mathcal{R})\,e^{-m_R\tilde{k}_\phi|\mathbf{x}|}-
\frac{4\,F(m_Y\mathcal{R})}{3}\,e^{-m_Y|\mathbf{x}|}\biggr]~,
\\\\
\Psi_{\rm ball}(\mathbf{x})\,=\,-\frac{GM}{|\mathbf{x}|}\biggl[1-g(\xi,\eta)\,F(m_R\tilde{k}_R\mathcal{R})\,e^{-m_R\tilde{k}_R|\mathbf{x}|}-[\frac{1}{3}-g(\xi,\eta)]\,F(m_R\tilde{k}_\phi\mathcal{R})\,e^{-m_R\tilde{k}_\phi|\mathbf{x}|}-
\frac{2\,F(m_Y\mathcal{R})}{3}\,e^{-m_Y|\mathbf{x}|}\biggr]~.
\end{array}
\end{eqnarray}
For $\mathbf{\Omega}(r)\,=\,\mathbf{\Omega}_0$, the metric potential (\ref{vect_pot2}) reads

\begin{eqnarray}
\mathbf{A}(\mathbf{x})\,=\,-\frac{3MG}{2\pi\mathcal{R}^3}\mathbf{\Omega}_0\times\int d^3\mathbf{x}'\frac{1-e^{-m_Y|\mathbf{x}-\mathbf{x}'|}}{|\mathbf{x}-\mathbf{x}'|}\,\Theta(\mathcal{R}-r')\,\mathbf{x}'~.
\end{eqnarray}
Making the approximation

\begin{eqnarray}
\frac{e^{-m_Y|\mathbf{x}-\mathbf{x}'|}}{|\mathbf{x}-\mathbf{x}'|}\,\sim\,\frac{e^{-m_Y\,r}}{r}+\frac{e^{-m_Y\,r}(1+m_Y\,r)\cos\alpha}{r}\frac{r'}{r}+\mathcal{O}\biggl(\frac{{r'}^2}{{r}^2}\biggr)~,
\end{eqnarray}
where $\alpha$ is the angle between the vectors $\mathbf{x}$, $\mathbf{x}'$, with $\mathbf{x}\,=\,r\,\hat{\mathbf{x}}$ where $\hat{\mathbf{x}}\,=\,(\sin\theta\cos\phi,\sin\theta\sin\phi,\cos\theta)$ and considering only the first order of $r'/r$, we can evaluate the integration in the vacuum ($r\,>\,\mathcal{R}$) as 

\begin{eqnarray}
\int d^3\mathbf{x}'\,\frac{e^{-m_Y|\mathbf{x}-\mathbf{x}'|}}{|\mathbf{x}-\mathbf{x}'|}\,\Theta(\mathcal{R}-r')\,\mathbf{x}'\,=\,\frac{4\pi}{15}\,\frac{(1+m_Y\,r)\,e^{-m_Y\,r}\,\mathcal{R}^5}{\,{r}^3}\,\mathbf{x}~.
\end{eqnarray}
Thus, the field $\mathbf{A}$ outside the sphere is

\begin{eqnarray}\label{sol_A_pointlike}
\mathbf{A}(\mathbf{x})\,=\,\frac{G}{|\mathbf{x}|^2}\,\biggl[1-(1+m_Y|\mathbf{x}|)\,e^{-m_Y|\mathbf{x}|}\biggr]\,\hat{\mathbf{x}}\times\mathbf{J}~,
\end{eqnarray}
where $\mathbf{J}\,=\,2M\mathcal{R}^2\mathbf{\Omega}_0/5$ is the angular momentum of the ball.

The modification with respect to  General Relativity has the same feature as the one generated by the point-like source \cite{stabstab}. From the definition of $m_R$ and $m_Y$ (\ref{mass_definition}),  we  note that the presence of a Ricci scalar function ($f_{RR}(0)\,\neq\,0$) appears only in $m_R$. Considering only  $f(R)$-gravity ($m_Y\,\rightarrow\,\infty$), the solution (\ref{sol_A_pointlike}) is unaffected by the modification in the Hilbert-Einstein action.

In the following, we will apply the above analysis in the case of bodies moving in the gravitational field.

\section{ The body motion  in the weak gravitational field}

Let us consider the  geodesic equations

\begin{eqnarray}\label{geodesic}
\frac{d^2\,x^\mu}{ds^2}+\Gamma^\mu_{\alpha\beta}\frac{dx^\alpha}{ds}\frac{dx^\beta}{ds}\,=\,0~,
\end{eqnarray}
where $ds\,=\,\sqrt{g_{\alpha\beta}dx^\alpha dx^\beta}$ is the relativistic distance. In terms of the potentials generated by the ball source with radius $\mathcal{R}$,  the components of the metric $g_{\mu\nu}$ read

\begin{eqnarray}
\nonumber
&&g_{tt}\,=\,1+2\Phi_{ball}(\mathbf{x})\,=\,1-\frac{2GM}{|\mathbf{x}|}\biggl[1+g(\xi,\eta)\,F(m_R\tilde{k}_R\mathcal{R})\,e^{-m_R\tilde{k}_R|\mathbf{x}|}+[1/3-g(\xi,\eta)]\,F(m_R\tilde{k}_\phi\mathcal{R})\,e^{-m_R\tilde{k}_\phi|\mathbf{x}|}\\	
\nonumber
&&\qquad\qquad\qquad\qquad\qquad\qquad\qquad\qquad-\frac{4\,F(m_Y\mathcal{R})}{3}\,e^{-m_Y|\mathbf{x}|}\biggr]~,\\
&&g_{ti}\,=\,2A_i(\mathbf{x})\,=\,\frac{2G}{|\mathbf{x}|^2}\,\biggl[1-(1+m_Y|\mathbf{x}|)\,e^{-m_Y|\mathbf{x}|}\biggr]\,\hat{\mathbf{x}}\times\mathbf{J}~,\\
\nonumber
&&g_{ij}\,=\,-\delta_{ij}+2\Psi_{ball}(\mathbf{x})\delta_{ij}\,=\,-\delta_{ij}\,-\frac{2GM}{|\mathbf{x}|}\biggl[1-g(\xi,\eta)\,F(m_R\tilde{k}_R\mathcal{R})\,e^{-m_R\tilde{k}_R|\mathbf{x}|}\\
\nonumber
&&\qquad\qquad\qquad\qquad\qquad\qquad\qquad\qquad\qquad\qquad-[1/3-g(\xi,\eta)]\,F(m_R\tilde{k}_\phi\mathcal{R})\,e^{-m_R\tilde{k}_\phi|\mathbf{x}|}-\frac{2\,F(m_Y\mathcal{R})}{3}\,e^{-m_Y|\mathbf{x}|}\biggr]\delta_{ij}~,
\end{eqnarray}
and the non-vanishing Christoffel symbols read

\begin{eqnarray}
\label{Christoffel}
	&&\Gamma^t_{ti}\,=\,\Gamma^i_{tt}=\partial_i
    \Phi_{\rm ball}~,\nonumber\\ &&\Gamma^i_{tj}\,=\,\frac{\partial_i A_j-\partial_j
    A_i}{2}~,\\ &&\Gamma^i_{jk}\,=\,\delta_{jk}\partial_i \Psi_{\rm ball}
    -\delta_{ij}\partial_k \Psi_{ball}-\delta_{ik}\partial_j \Psi_{\rm ball}~.\nonumber
\end{eqnarray}
Let us consider some specific motions.

\subsection{Circular rotation curves in a spherically symmetric field}\label{CRCSTFOG_par}

In the Newtonian limit, Eq.(\ref{geodesic}), neglecting the rotating component of the source, leads to the usual equation of motion of bodies

\begin{eqnarray}
\frac{d^2\,\mathbf{x}}{dt^2}\,=\,-\nabla\Phi_{\rm ball}(\mathbf{x})~,
\end{eqnarray}
where the gravitational potential is given by Eq.~(\ref{ST_FOG_FE_NL_sol_ball}). The study of motion is very simple  considering a particular symmetry for mass distribution $\rho$, otherwise  analytical solutions are not available. However, our aim is to evaluate the corrections to the classical motion in the easiest situation, namely the circular motion, in which case  we do not consider  radial and vertical motions. The condition of stationary motion on the circular orbit reads

\begin{eqnarray}\label{stazionary_motion}
v_{\rm c}(r)\,=\,\sqrt{r\,\frac{\partial\Phi(r)}{\partial r}}~,
\end{eqnarray}
where $v_{\rm c}$ denotes the velocity.

A further remark on Eq. (\ref{ST_FOG_FE_NL_sol_point}) is needed. The structure of solutions is mathematically similar to the one of fourth-order gravity $f(R,R_{\alpha\beta}R^{\alpha\beta})$, however  there is a fundamental difference regarding the algebraic signs of the Yukawa corrections. More precisely, whilst the Yukawa correction induced by a generic function of the Ricci scalar leads to an attractive  gravitational force, and the one induced by Ricci tensor squared leads to a repulsive one~\cite{stabscel}, here the Yukawa corrections induced by a generic function of Ricci scalar and  a nonminimally coupled scalar field, have both a positive coefficient (see for details Ref.~\cite{FOGST}). Hence the scalar field gives rise to  a stronger attractive force than in  $f(R)$-gravity, which may imply that $f(R,\phi)$-gravity is a better choice than  $f(R,R_{\alpha\beta}R^{\alpha\beta})$-gravity. However, there is a problem in the limit $|\textbf{x}|\,\rightarrow\,\infty$: the interaction is scale-depended (the scalar fields are massive) and, in the vacuum, the corrections turn off. Thus,  at large distances, we recover only the classical Newtonian contribution. In conclusion, the presence of scalar fields makes the  profile smooth, a behavior which is apparent in the study of rotation curves.

For an illustration, let us consider the  phenomenological potential $\Phi_{\rm SP}(r)\,=\,-\frac{GM}{r}\biggl[1+\alpha\,e^{-m_{\rm S}\,r}\biggr]$, with $\alpha$ and $m_{\rm S}$ free parameters, chosen by Sanders~\cite{sanders} in an attempt to fit galactic rotation curves of spiral galaxies in the absence of dark matter, within the MOdified Newtonian
Dynamics (MOND) proposal of Milgrom~\cite{Milgrom:1983ca}, was further accompanied by a relativistic partner known as Tensor-Vector-Scalar (TeVES) model~\cite{Bekenstein:2004ne}\footnote{Note that the validity of MOND~\cite{Ferreras:2007kw} and TeVeS~\cite{Mavromatos:2009xh,Ferreras:2009rv,Ferreras:2012fg} models of
modified gravity were tested by using gravitational lensing
techniques, with the conclusion that a non-trivial component in the
form of dark matter has to be added to those models in order to match
the observations.  However, there are proposals of
modified gravity, as for instance the string inspired model studied in
Ref.~\cite{Mavromatos:2007sp}, leading to an
action that includes, apart from the metric tensor field, also scalar
(dilaton) and vector fields, which may be in agreement with current
observational data. Note that this model, based on brane universes propagating in
bulk space-times populated by point-like defects does have
dark matter components, while the r\^ole
of extra dark matter is also provided by the population of massive defects~\cite{Mavromatos:2012ha}.}. The free parameters selected by Sanders were  $\alpha\,\simeq\,-0.92$ and $1/m_{\rm S}\,\simeq\,40\, \text{Kpc}$.   Note that this potential were recently used for elliptical galaxies~\cite{capdena}. 
In both cases,  assuming  a negative value for $\alpha$, an almost  constant profile for rotation curve is recovered, however there are two issues. Firstly, an $f(R, \phi)$-gravity does not lead to  that negative value of $\alpha$, and secondly the presence of Yukawa-like correction with negative coefficient leads to a lower rotation curve and only by resetting $G$ one  can fit the experimental data.

Only if we consider a massive, non minimally coupled scalar-tensor theory,   we get a potential with negative coefficient in Eq.~(\ref{ST_FOG_FE_NL_sol_point})~\cite{FOGST}. In fact  setting the gravitational constant equal to ${\displaystyle G_0\,=\,\frac{2\,\omega(\phi^{(0)})\,\phi^{(0)}-4}{2\,\omega(\phi^{(0)})\,\phi^{(0)}-3}\frac{G_\infty}{\phi^{(0)}}}$, where $G_\infty$ is the gravitational constant as measured at infinity, and imposing $\alpha^{-1}\,=\,3-2\,\omega(\phi^{(0)})\,\phi^{(0)}$, the potential (\ref{ST_FOG_FE_NL_sol_point}) becomes $\Phi(r)\,=\,-\frac{G_\infty M}{r}\biggl\{1+\alpha\,e^{-\sqrt{1-3\alpha}\,m_\phi r}\biggr\}$ and then the Sanders potential can be recovered.

\begin{figure}[htbp]
\centering
\includegraphics[scale=1]{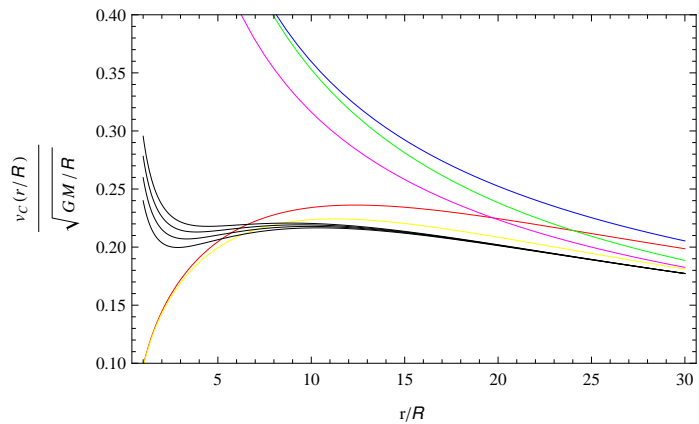}
\caption{The circular velocity of a ball source of mass $M$ and radius $\mathcal{R}$, with the potentials of Table \ref{table_potential}. We indicate case A by green line, case B by yellow line, case D by  red line, case C by  blue line, and the GR case by magenta line. The black line correspond to the Sanders model for $-0.95\,<\,\alpha\,<\,-0.92$. The values of free parameters are: $\omega(\phi^{(0)})\,=\,-1/2$, $\xi\,=\,-5$, $\eta\,=\,.3$, $m_Y\,=\,1.5*m_R$, $m_S\,=\,1.5*m_R$, $m_R\,=\,.1*\mathcal{R}^{-1}$}
\label{fig_1}
\end{figure}

In Fig.~\ref{fig_1} we show the radial behaviour of the circular velocity induced by the presence of a ball source in the case of the Sanders potential and of potentials shown in Table \ref{table_potential}.

\begin{table}[ht]
\centering
\begin{tabular}{c|c|c|c}
\hline\hline\hline
Case & Theory & Gravitational potential & Free parameters \\
\hline
& & & \\
A & $f(R)$ & $-\frac{GM}{|\mathbf{x}|}\biggl[1+\frac{1}{3}\,e^{-m_R|\mathbf{x}|}\biggl]$ & $\begin{array}{ll}{m_R}^2\,=\,-\frac{1}{3f_{RR}(0)}\end{array}$ \\
\hline
& & & \\
B & $f(R,\,R_{\alpha\beta}R^{\alpha\beta})$ & $-\frac{GM}{|\mathbf{x}|}\biggl[1+\frac{1}{3}\,e^{-m_R|\mathbf{x}|}-\frac{4}{3}\,e^{-m_Y|\mathbf{x}|}\biggl]$ & $\begin{array}{ll}{m_R}^2\,=\,-\frac{1}{3f_{RR}(0,0)+2f_Y(0,0)}\\\\{m_Y}^2\,=\,\frac{1}{f_Y(0,0)}
\end{array}$ \\
\hline
&  & & \\
C & $f(R,\,\phi)+\omega(\phi)\phi_{;\alpha}\phi^{;\alpha}$ & $\begin{array}{ll}-\frac{GM}{|\mathbf{x}|}\biggl[1+g(\xi,\eta)\,e^{-m_R\tilde{k}_R\,|\textbf{x}|}\\\\\qquad\qquad+[1/3-g(\xi,\eta)]\,e^{-m_R\tilde{k}_\phi\,|\textbf{x}|}\biggr]\end{array}$ & $\begin{array}{ll}{m_R}^2\,=\,-\frac{1}{3f_{RR}(0,\phi^{(0)})}\\\\{m_\phi}^2\,=\,-\frac{f_{\phi\phi}(0,\phi^{(0)})}{2\omega(\phi^{(0)})}\\\\\xi\,=\,\frac{3{f_{R\phi}(0,\phi^{(0)})}^2}{2\omega(\phi^{(0)})}\\\\\eta\,=\,\frac{m_\phi}{m_R}\\\\g(\xi,\,\eta)\,=\,\frac{1-\eta^2+\xi+\sqrt{\eta^4+(\xi-1)^2-2\eta^2(\xi+1)}}{6\sqrt{\eta^4+(\xi-1)^2-2\eta^2(\xi+1)}}\\\\{\tilde{k}_{R,\phi}}^2\,=\,\frac{1-\xi+\eta^2\pm\sqrt{(1-\xi+\eta^2)^2-4\eta^2}}{2}
\end{array}$ \\
\hline
&  & & \\
D & $f(R,\,R_{\alpha\beta}R^{\alpha\beta},\phi)+\omega(\phi)\phi_{;\alpha}\phi^{;\alpha}$ & $\begin{array}{ll}-\frac{GM}{|\mathbf{x}|}\biggl[1+g(\xi,\eta)\,e^{-m_R\tilde{k}_R\,|\textbf{x}|}\\\\\,\,\,\,+[1/3-g(\xi,\eta)]\,e^{-m_R\tilde{k}_\phi\,|\textbf{x}|}-\frac{4}{3}\,e^{-m_Y|\mathbf{x}|}\biggr]\end{array}$ & $\begin{array}{ll}{m_R}^2\,=\,-\frac{1}{3f_{RR}(0,0,\phi^{(0)})+2f_Y(0,0,\phi^{(0)})}\\\\{m_Y}^2\,=\,\frac{1}{f_Y(0,0,\phi^{(0)})}\\\\{m_\phi}^2\,=\,-\frac{f_{\phi\phi}(0,0,\phi^{(0)})}{2\omega(\phi^{(0)})}\\\\\xi\,=\,\frac{3{f_{R\phi}(0,0,\phi^{(0)})}^2}{2\omega(\phi^{(0)})}\\\\\eta\,=\,\frac{m_\phi}{m_R}\\\\g(\xi,\,\eta)\,=\,\frac{1-\eta^2+\xi+\sqrt{\eta^4+(\xi-1)^2-2\eta^2(\xi+1)}}{6\sqrt{\eta^4+(\xi-1)^2-2\eta^2(\xi+1)}}\\\\{\tilde{k}_{R,\phi}}^2\,=\,\frac{1-\xi+\eta^2\pm\sqrt{(1-\xi+\eta^2)^2-4\eta^2}}{2}
\end{array}$ \\
\hline\hline\hline
\end{tabular}
\caption{\label{table_potential}Table of  fourth order gravity models analyzed in the Newtonian limit for gravitational potentials generated by a point-like source Eq.~(\ref{ST_FOG_FE_NL_sol_point}). The range of validity of cases C, D is $(\eta-1)^2-\xi\,>\,0$. We set $f_R(0,0,\phi^{(0)})\,=\,1$.}
\end{table}

\subsection{Rotating sources and orbital parameters}\label{RSPSTFOG_par}

Considering the geodesic equations (\ref{geodesic}) with the Christoffel symbols given in Eq.~(\ref{Christoffel}), we obtain
\begin{eqnarray}
\frac{d^2x^i}{ds^2}+\Gamma^{i}_{tt}+2\Gamma^{i}_{tj}\frac{dx^j}{ds}\,=\,0~,
\end{eqnarray}
which in the coordinate system ${\bf J}\,=\,(0,0,J)$, reads

\begin{eqnarray}\label{GeoEq2}
\nonumber
&&\ddot{x}+\frac{GM}{r^3}x\,=\,-\frac{GM\Lambda(r)}{r^3}x+\frac{2GJ}{r^5} \biggl\lbrace \zeta( r) \biggl[ \biggl( x^2+y^2-2z^2\biggl)\dot{y}+3yz\dot{z}\biggr]+2\Sigma(r)L_x z \biggr\rbrace~,\\
&&\ddot{y}+\frac{GM}{r^3}y\,=\,-\frac{GM\Lambda(r)}{r^3}y-\frac{2GJ}{r^5} \biggl\lbrace \zeta( r) \biggl[ \biggl( x^2+y^2-2z^2\biggl)\dot{x}+3xz\dot{z}\biggl]-2\Sigma(r)L_y z \biggr\rbrace~,	\\
\nonumber
&&\ddot{z}+\frac{GM}{r^3}z\,=\,-\frac{GM\Lambda(r)}{r^3}z+\frac{6GJ}{r^5} \biggl\lbrace \zeta( r)+\frac{2}{3}\Sigma(r) \biggr\rbrace L_z z~,
\end{eqnarray}
where

\begin{eqnarray}
\nonumber
&&\Lambda(r)\,\doteq\,g(\xi,\eta)\,F(m_R\tilde{k}_R\mathcal{R})\,(1+m_R\tilde{k}_R r)\,e^{-m_R\tilde{k}_R r}+[1/3-g(\xi,\eta)]\,F(m_R\tilde{k}_\phi\mathcal{R})\,(1+m_R\tilde{k}_\phi r)\,e^{-m_R\tilde{k}_\phi r}\\&&\qquad\qquad-\frac{4\,F(m_Y\mathcal{R})}{3}\,(1+m_Y r)\,e^{-m_Y r}~,\nonumber
\\
&&\zeta( r)\,\doteq\,1-\bigl[1+m_Y r+(m_Y r)^2\bigr]e^{-m_Y r}~,\\
\nonumber
&&\Sigma(r)\,\doteq\,(m_Y r)^2e^{-m_Y r}\,~,
\end{eqnarray}
with $L_x$,$L_y$ and $L_z$ the components of the angular momentum.

The first terms in the right-hand-side of Eq.~(\ref{GeoEq2}), depending on the three parameters $m_R,m_Y$ and $m_\phi$, represent the Extended Gravity (EG) modification of the Newtonian acceleration. The second terms in these equations, depending on the angular momentum $J$ and the EG parameters $m_R,m_Y$ and $m_\phi$, correspond to dragging contributions. The case $m_R\rightarrow\infty,\,m_Y\rightarrow\infty$ and $m_\phi\rightarrow 0$  leads to $\Lambda(r)\rightarrow 0$, $\zeta( r)\rightarrow 1$  and $\Sigma(r)\rightarrow 0$, and hence one recovers the familiar results of GR ~\cite{LenseThirring}. These additional gravitational terms can be considered as perturbations of Newtonian gravity, and their effects on planetary motions can be calculated within the usual perturbative schemes assuming the Gauss equations~\cite{RoyOrbit}. We will follow this approach in what follows.

\begin{figure}[htbp]
\centering
\includegraphics[width=15cm]{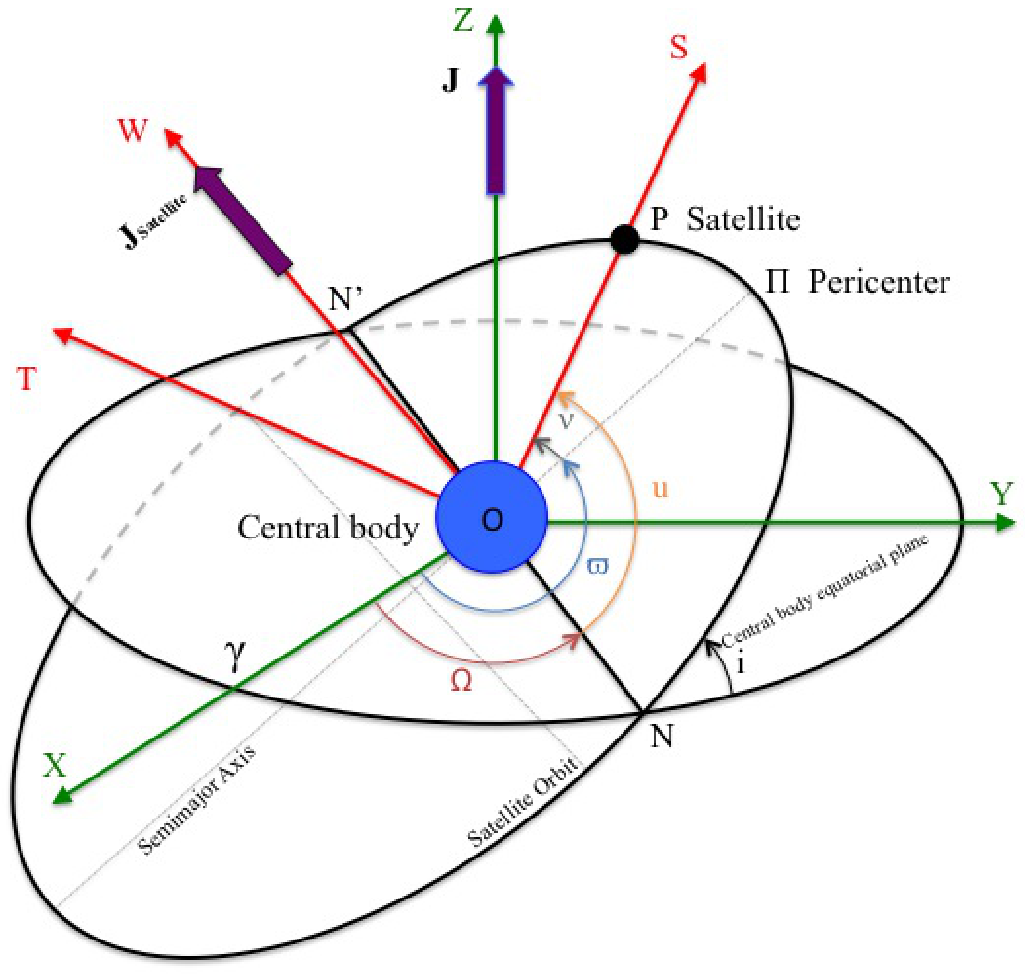}
\caption{$i$\,=\,\,$)\hspace{-9pt}<$ $YN\Pi$ is the inclination; $\Omega$\,=\,\,$)\hspace{-9.3pt}<$ $XON$ is the longitude of the ascending node $N$; $\tilde{\omega}$= broken $)\hspace{-8.8pt}<$ $XO\Pi$ is the longitude of the pericenter $\Pi$;  $\nu$\,=\,\,$)\hspace{-9pt}<$ $\Pi OP$ is the true anomaly; $u$\,=\,\,$)\hspace{-9pt}<$ $\Omega OP$=$\nu+\tilde{\omega}-\Omega$ is the argument of the latitude;  $\textbf{J}$ is the angular momentum of rotation of the central body; and $\textbf{J}_{\rm Satellite}$ is the angular momentum of revolution of a satellite around the central body.}
\label{Orbit}
\end{figure}

Let us consider the right-hand-side of Eq.~(\ref{GeoEq2}) as the components $(A_x,A_y,A_z)$ of the perturbing acceleration in the system ($X, Y, Z$) (see Fig.~\ref{Orbit}), with $X$ the axis passing through the vernal equinox $\gamma$, $Y$ the transversal axis, and $Z$ the orthogonal axis parallel to the angular momentum $\textbf{J}$ of the central body. In the system ($S, T, W$), the three components can be expressed as $(A_s,A_t,A_w)$, with $S$ the radial axis, $T$ the transversal axis, and $W$ the orthogonal one. We will adopt the standard notation: $a$ is the semimajor axis; $e$ is the eccentricity; $p=a(1-e^2)$ is the semilatus rectum; $i$ is the inclination; $\Omega$ is the longitude of the ascending node $N$; $\tilde{\omega}$ is the longitude of the pericenter $\Pi$; $M^0$ is the longitude of the satellite at time $t=0$; $\nu$ is the true anomaly; $u$ is the argument of the latitude given by $u=\nu+\tilde{\omega}-\Omega$; $n$ is the mean daily motion equal to $n=(GM/a^3)^{1/2}$; and $C$ is twice the velocity, namely $C=r^2\dot{\nu}a^2(1-e^2)^{1/2}$.

The transformation rules between the coordinates frames ($X, Y, Z$) and ($S, T, W$) are

\begin{eqnarray}
\begin{array}{ll}
x\,=\,r(\cos u\cos\Omega - \sin u\sin\Omega\cos i)~,\\
y\,=\,r(\cos u\sin\Omega + \sin u\cos\Omega\cos i)~,\\
z\,=\,r\sin u\sin i\\
r\,=\,\frac{p}{1+e\cos\nu}~,
\end{array}
\end{eqnarray}
and the components of the angular momentum obey the equations

\begin{eqnarray}
\nonumber
&&L_x\,=\,y\dot{z}-z\dot{y}=C\sin i\,\sin\Omega~,\\
&&L_y\,=\,z\dot{x}-x\dot{z}=-C\cos\Omega\,\sin i~,\\
\nonumber
&&L_z\,=\,x\dot{y}-y\dot{x}=C\cos i~.
\end{eqnarray}

The components of the perturbing acceleration in the ($S,\,T,\,W$) system read
\begin{eqnarray}\label{ForseSTW}
\nonumber
&&A_s\,=\,-\frac{G\,M\,\Lambda(r)}{r^2}+\frac{2\,G\,J\, C\cos i}{r^4} \zeta( r)~,\\
&&A_t\,=\,-\frac{2\,G\,J\, C\,e\cos\,i\sin\nu}{p\,r^3} \zeta( r)~,\\
\nonumber
&&A_w\,=\,\frac{2\,G\,J\, C\sin i}{r^4}\left[ \left(\frac{r\,e\sin\nu\cos u}{p}+2\sin u \right)\zeta( r) +2\sin u\,\Sigma(r)\right]~.
\end{eqnarray}
The $A_s$ component has two contributions: the former one results from the modified Newtonian potential $\Phi_{\rm ball}(\mathbf{x})$, while the latter one results from the gravito-magnetic field $A_i$ and it is a higher order term than the first one. Note that the components $A_t$ and $A_w$ depend only on the gravito-magnetic field.
The Gauss equations for the variations of the six orbital parameters, resulting from the perturbing acceleration with components $A_x, A_y, A_z$, read

\begin{eqnarray}\label{OrbitalParameters}
\begin{array}{ll}
\frac{da}{dt}\,= \,\dot{a}_{\rm EG}\,=\,\frac{2e\,G\,M\,\Lambda(r)\,\sin \nu}{n\,\,\sqrt[]{1-e^2}\,C}\dot{\nu}~,
\\
\frac{de}{dt}\,=\,\dot{e}_{\rm GR}+ \dot{e}_{\rm EG}\,=\,\frac{\sqrt[]{1-e^2}\,G\,M\,\Lambda(r)\,\sin \nu}{n\,a\,C}\dot{\nu}+\dot{e}_{\rm GR}\biggl[1-\,e^{-m_Y r}\Bigl(1+m_Y r+(m_Y r)^2\Bigr)\biggl]~,
\\
\frac{d\Omega}{dt}\,=\,\dot{\Omega}_{\rm GR}+ \dot{\Omega}_{\rm EG}\,=\,\dot{\Omega}_{\rm GR}\biggl\{1-\,e^{-m_Y r}\Bigl[  1+m_Y r+\bigl(1+f(\nu,u,e)\bigl) (m_Y r)^2 \Bigl]\biggl\}~,
\\
\frac{di}{dt}\,=\,\dot{i}_{\rm GR}+ \dot{i}_{\rm EG}\,=\,\dot{i}_{\rm GR}\biggl\{1-\,e^{-m_Y r}\Bigl[  1+m_Y r+\bigl(1+f(\nu,u,e)\bigl) (m_Y r)^2 \Bigl]\biggl\}~,
\\
\frac{d\tilde{\omega}}{dt}\,=\,\dot{\tilde{\omega}}_{\rm GR}+ \dot{\tilde{\omega}}_{\rm EG}\,=\,-\frac{\sqrt[]{1-e^2}\,G\,M\,\Lambda(r)\,\cos \nu}{n\,a\,e\,C}\dot{\nu}+\dot{\tilde{\omega}}_{\rm GR}
\biggl[1-\,e^{-m_Y r}\Bigl(1+m_Y r+(m_Y r)^2\Bigr)\biggl]
\\\qquad\qquad\qquad\qquad\qquad\qquad\qquad\qquad\qquad\qquad\qquad\qquad\qquad-2\sin^2\frac{i}{2}\,\dot{\Omega}_{\rm GR}f(\nu,u,e)\Sigma( r)~,
\\
\frac{dM^0}{dt}\,=\,\dot{M^0}_{\rm GR}+ \dot{M^0}_{\rm EG}\,=\,-\frac{G\,M\,\Lambda(r)}{n\,a\,C}\biggl[\frac{2r}{a}+\frac{e\,\,\,\sqrt[]{1-e^2}}{1+\,\,\sqrt[]{1-e^2}}\cos \nu  \biggl]\dot{\nu}+\dot{M^0}_{\rm GR}\biggl[1-\,e^{-m_Y r}\Bigl(1+m_Y r+(m_Y r)^2\Bigr)\biggl]
\\\qquad\qquad\qquad\qquad\qquad\qquad\qquad\qquad\qquad\qquad\qquad\qquad\qquad\qquad\qquad\qquad\qquad-2\sin^2\frac{i}{2}\,\dot{\Omega}_{\rm GR}f(\nu,u,e)\Sigma(r)~,
\end{array}
\end{eqnarray}
where

\begin{eqnarray}
\begin{array}{ll}
\dot{e}_{\rm GR}\,=\,\frac{2\,G\,J\cos i\sin\nu}{a\,C}\dot{\nu}~,
\\
\dot{\Omega}_{\rm GR}\,=\,\frac{2\,G\,J\sin u}{p\,C}\biggl[ e\sin\nu\cos u+2\biggl( 1+e\cos\nu\biggl)\sin u \biggl] \dot{\nu}~,
\\
\dot{i}_{\rm GR}\,=\,\frac{2\,G\,J\cos u\sin i}{C\,p}\biggl[ e\sin\nu\cos u+2\biggl( 1+e\cos\nu\biggl)\sin u \biggl] \dot{\nu}~,
\\
\dot{\tilde{\omega}}_{\rm GR}\,=\,-\frac{2\,G\,J\cos i}{a\, C}\left( 2+\frac{1+e^2}{e}\cos\nu\right)\dot{\nu} +2\sin^2 \frac{i}{2}\,\dot{\Omega}_{\rm GR}~,
\\
\dot{{\cal M}^0}_{\rm GR}\,=\,-\frac{4\,G\,J\cos i}{n\,a^2\,p}\biggl( 1+e\cos\nu\biggl)\dot{\nu}+\frac{e^2}{1+\,\,\sqrt[]{1-e^2} }\dot{\tilde{\omega}}_{\rm GR}+2\,\,\sqrt[]{1-e^2} \sin^2 \frac{i}{2}\,\dot{\Omega}_{\rm GR}~,
\\
f(\nu,u,e)\,=\,\frac{1+e\cos\nu}{1+e\left(\frac{\sin\nu\cot u}{2}+\cos\nu \right) }~.
\end{array}
\end{eqnarray}
Hence, we have derived the corresponding equations of the six orbital parameters for Extended Gravity, with the dynamics of $a, e, \tilde{\omega}, L^0$ depending mainly on the terms related to the modifications of the Newtonian potential, whilst the dynamics of $\Omega$ and $i$ depending only on the dragging terms.

Considering an almost circular orbit ($e\ll 1$), we integrate the Gauss equations with respect to the only anomaly $\nu$, from $0$ to $\nu(t)\,=\,nt$, since all other parameters have a slower evolution than $\nu$, hence they can be considered as constraints with respect to $\nu$. At first order we get

\begin{eqnarray}\label{par_sol}
\begin{array}{ll}
\Delta a(t)\,=\,0~,
\\
\Delta e(t)\,=\,0~,
\\
\Delta i(t)\,=\,\frac{G\,J\,e^2\,\sin i}{ na^3}e^{-m_Y p }(m_Y p)^2\biggl[ 1+\frac{(m_Y p)^2}{2}\biggl(m_Y p-4 \biggl) \biggl]\sin\biggl(\tilde{\omega}(t)-\Omega(t) \biggl)\nu(t)+{\cal O}(e^4)~,
\\
\Delta \Omega(t)\,=\,\frac{2\,G\,J}{ na^3}\biggl[ 1-e^{-m_Y p }\biggl(  1+m_Y p+ 2(m_Y p)^2\biggl)  \biggl]\nu(t)+ {\cal O}(e^2)~,
\\
\Delta\tilde{\omega}(t)\,=\,\biggl\lbrace \frac{\tilde{\Lambda}(p)}{2}-\frac{2\,G\,J}{na^3}\biggl[ 3\cos i -1+e^{-m_Y p }( 1+m_Y p +\frac{3}{2}(m_Y p)^2
\\
\qquad\qquad\qquad\qquad\qquad\qquad -( 3+3m_Y p +3(m_Y p)^2+\frac{1}{12}(m_Y p)^3)\cos i)\biggl]\biggr\rbrace \nu(t)+ {\cal O}(e^2)~,
\\
\Delta {\cal M}^0(t)\,=\,\biggl\lbrace 2\Lambda(p)-\frac{2\,G\,J}{ n a^3}\biggl[ 3\cos i-1-e^{-m_Y p }\biggl(  1+m_Y p+ 2(m_Y p)^2\biggl) \biggl( \cos i -1\biggl)\biggl]\biggr\rbrace \nu(t)+ {\cal O}(e^2)~,
\end{array}
\end{eqnarray}
where

\begin{eqnarray}
\tilde{\Lambda}(p)&\doteq &g(\xi,\eta)\,F(m_R\tilde{k}_R\mathcal{R})\,(m_R\tilde{k}_R p)^2\,e^{-m_R\tilde{k}_R p}+[1/3-g(\xi,\eta)]\,F(m_R\tilde{k}_\phi\mathcal{R})\,(m_R\tilde{k}_\phi p)^2\,e^{-m_R\tilde{k}_\phi p}\nonumber\\\\
&&-\frac{4\,F(m_Y\mathcal{R})}{3}\,(m_Y p)^2\,e^{-m_Y p}~.\nonumber
\end{eqnarray}

We hence notice that the contributions to the semimajor axis $a$ and eccentricity $e$ vanish, as in GR, whilst there are nonzero contributions to $i$, $\Omega$, $\tilde{\omega}$ and $M^0$. In particular,
the contributions to the inclination $i$ and the longitude of the ascending node $\Omega$,  depend only on the drag effects of the rotating central body; while the contributions to the pericenter longitude  $\tilde{\omega}$ and mean longitude at $M^0$, depend also on the modified Newtonian potential. Finally, note that in the Extended Gravity model we have considered here, the inclination $i$ has a nonzero contribution, in contrast to the result obtained within GR, and also $\Delta\tilde{\omega}(t)\neq\Delta M^0(t)$, given by

\begin{eqnarray}
\Delta\tilde{\omega}(t)-\Delta {\cal M}^0(t)\simeq && \biggl\{\frac{\tilde{\Lambda}(p)-4\Lambda(p)}{2} +\frac{2GJ}{na^3}e^{-m_Y p }\biggl[\frac{(m_Y p)^2}{2}+( 2+2m_Y p+ (m_Y p)^2\nonumber\\\\
&&\qquad+\frac{(m_Y p)^3}{12})\cos i\biggr] \biggl\} \,\nu(t)+ {\cal O}(e^2)~.\nonumber
\end{eqnarray}
In the limit $m_R\rightarrow\infty, m_Y\rightarrow\infty$ and $m_\phi\rightarrow 0$, we obtain the well-known  results of GR.

\section{Experimental constraints}\label{EXSTFOG_par}

The orbiting gyroscope precession can be split into a part generated by the metric potentials, $\Phi$ and $\Psi$, and one generated by the vector potential $\mathbf{A}$. The equation of motion for the gyro-spin three-vector $\mathbf{S}$ is

\begin{equation}\label{dSdt}
\frac{d\mathbf{S}}{dt}\,=\,\frac{d\mathbf{S}}{dt}\Big|_\text{G}+\frac{d\mathbf{S}}{dt}\Big|_\text{LT}
\end{equation}
where the geodesic and Lense-Thirring precessions are

\begin{eqnarray}\label{dS-G}
&&\frac{d\mathbf{S}}{dt}\Big|_\text{G}\,=\,\mathbf{\Omega}_\text{G}\times\mathbf{S}\,\,\,\,\text{with}\,\,\,\,\mathbf{\Omega}_\text{G}\,=\,\frac{\nabla(\Phi+2\Psi)}{2}\times\mathbf{v}~,
\nonumber\\\\
&&\frac{d\mathbf{S}}{dt}\Big|_\text{LT}\,=\,\mathbf{\Omega}_\text{LT}\times\mathbf{S}\,\,\,\,\text{with}\,\,\,\,\mathbf{\Omega}_\text{LT}\,=\,\frac{\nabla\times\mathbf{A}}{2}~.\nonumber
\end{eqnarray}
The geodesic precession, $\mathbf{\Omega}_G$, can be written as the sum of two terms, one obtained with GR and the other being the Extended Gravity contribution. Then we have

\begin{equation}\label{OmegaGfinal}
\mathbf{\Omega}_\text{G}\,=\,\mathbf{\Omega}_\text{G}^\text{(GR)}+\mathbf{\Omega}_\text{G}^\text{(EG)}~,
\end{equation}
where

\begin{eqnarray}\label{OmegaGE}
&&\mathbf{\Omega}_\text{G}^\text{(GR)}\,=\,\frac{3GM}{2|\mathbf{x}|^3}\,\,\mathbf{x}\times\mathbf{v}~,
\nonumber\\\nonumber\\
&&\mathbf{\Omega}_\text{G}^\text{(EG)}\,=\,-\biggl[g(\xi,\eta)(m_R\tilde{k}_Rr+1)\,F(m_R\tilde{k}_R\mathcal{R})\,e^{-m_R\tilde{k}_Rr}+
\frac{8}{3}(m_Yr+1)\,F(m_Y\mathcal{R})\,e^{-m_Yr}\\\nonumber
&&\qquad\qquad\qquad\qquad\qquad\qquad\qquad\qquad +\biggl[\frac{1}{3}-g(\xi,\eta)](m_R\tilde{k}_\phi r+1)F(m_R\tilde{k}_\phi\mathcal{R})\,e^{-m_R\tilde{k}_\phi r}\biggr]\frac{{\bf \Omega}_{\rm G }^{\rm (GR)}}{3}~.
\end{eqnarray}
where $|\textbf{x}|=r$. Similarly one has

\begin{eqnarray}
\label{OmegaLTfinal}
\mathbf{\Omega}_\text{LT}\,=\,\mathbf{\Omega}_\text{LT}^\text{(GR)}+\mathbf{\Omega}_\text{LT}^\text{(EG)}~,
\end{eqnarray}
with

\begin{eqnarray}
\label{OmegaLTaverage}
&&\mathbf{\Omega}_\text{LT}^\text{(GR)}\,=\,\frac{G}{2r^3}\,\mathbf{J}~,
\nonumber\\\\
&&\mathbf{\Omega}_\text{LT}^\text{(EG)}\,=\,-e^{-m_Y r}(1+m_Y r+{m_Y}^2r^2)\,\mathbf{\Omega}_\text{LT}^\text{(GR)}~,\nonumber
\end{eqnarray}
where we have assumed that, on the average, $\langle(\mathbf{J}\cdot\mathbf{x})\,\mathbf{x}\rangle\,=\,0$.

Gravity Probe B satellite contains a set of four gyroscopes and has tested two predictions of GR: the geodetic effect and frame-dragging (Lense-Thirring effect).   The tiny changes in the direction of spin gyroscopes, contained in the satellite orbiting at $h=650$ km of altitude and crossing directly over the poles, have been measured with extreme precision. The values of the geodesic precession and the Lense-Thirring precession, measured by the Gravity Probe B satellite and those predicted by GR, are given in Table II.
\begin{table}[t]
\caption{The geodesic precession and Lense-Thitting (frame dragging) precession as predicted by GR and observed with the Gravity Probe B experiment~\cite{GPBexp}. }
\begin{tabular}{ccc} \\ \hline\hline
Effect & \ \ \ \ \ \ Measured (mas/y) & \ \ \ \ \ \ Predicted (mas/y) \\
  \hline Geodesic precession &\ \ \ \ \ \ $6602\pm 18$ &\ \ \ \ \ \
  6606 \\ \hline Lense-Thirring precession & $\ \ \ \ \ \ 37.2\pm 7.2$
  &\ \ \ \ \ \ 39.2 \\ \hline\hline
\end{tabular}
\end{table}
Imposing the constraint $|\mathbf{\Omega}_\text{G}^\text{(EG)}|\,\lesssim\,\delta \mathbf{\Omega}_\text{G}$ and $|\mathbf{\Omega}_\text{LT}^\text{(EG)}|\,\lesssim\,\delta \mathbf{\Omega}_\text{LT}$, \cite{LamSakSta}, with $r^*\,=\,R_\oplus+h$ where $R_\oplus$ is the radius of  the Earth and $h\,=\,650\,km$ is the altitude of the satellite, we get

\begin{eqnarray}\label{bound}
&& g(\xi,\eta)\big(m_R\tilde{k}_R\,r^*+1)\,F(m_R\tilde{k}_R R_\oplus)\,e^{-m_R\tilde{k}_R\,r^*}+[1/3-g(\xi,\eta)](m_R\tilde{k}_\phi\,r^*+1)\,F(m_R\tilde{k}_\phi R_\oplus)\,e^{-m_R\tilde{k}_\phi\,r^*}
\nonumber\\\nonumber\\
&&\qquad\qquad\qquad\qquad\qquad\qquad\qquad\qquad\qquad+\frac{8}{3}(m_Y\,r^*+1)\,F(m_Y R_\oplus)\,e^{-m_Y\,r^*}\,\lesssim\,\frac{3\,\delta|\mathbf{\Omega}_\text{G}|}{|\mathbf{\Omega}_\text{G}^\text{(GR)}|}\simeq 0.008~,
\\\nonumber\\
&&(1+m_Y r^* +{m_Y}^2 {r^*}^2)\,e^{-m_Y r^*}\,\lesssim\,\frac{\,\delta|\mathbf{\Omega}_\text{LT}|}{|\mathbf{\Omega}_\text{LT}^\text{(GR)}|}\simeq 0.19~,\nonumber
\end{eqnarray}
since, from the experiments, we have $|\mathbf{\Omega}_\text{G}^\text{(GR)}|\,=\,6606\, {\rm mas}$ and $\delta|\mathbf{\Omega}_\text{G}|\,=\,18\,{\rm mas}$, $|\mathbf{\Omega}_\text{LT}^\text{(GR)}|\,=\,37.2\,{\rm mas}$ and $\delta|\mathbf{\Omega}_\text{LT}|\,=\,7.2\,{\rm mas}$. From Eq.~(\ref{bound}) we thus obtain that $m_Y\geq 7.3\times 10^{-7} m^{-1}$.

The LAser RElativity Satellite (LARES) mission~\cite{laresdata} of the Italian Space Agency is designed to test the frame-dragging and the Lense-Thirring effect, to within 1$\%$ of the value predicted in the framework of GR. The body of this satellite has a diameter of about 36.4~cm and weights about 400~kg. It was inserted in an orbit with 1450~km of perigee, an inclination of $69.5\pm 1$ degrees and eccentricity $9.54\times10^{-4}$. It allows us to obtain a stronger constraint for $m_Y$:

\begin{eqnarray}
\label{boundLT2}
(1+m_Y r^*+{m_Y}^2{r^*}^2)\,e^{-m_Y r^*}\,\lesssim\,\frac{\,\delta|\mathbf{\Omega}_\text{LT}|}{|\mathbf{\Omega}_\text{LT}^\text{(GR)}|}\simeq 0.01~,
\end{eqnarray}
from the which we obtain $m_Y\geq 1.2\times 10^{-6} m^{-1}$.

In the specific case of the Noncommutative Spectral Geometry model, the quantities (\ref{mass_definition}) become $m_R\rightarrow \infty$, $m_Y=\sqrt{\frac{5\pi^2\bigl(k_0^2{\bf H}^{(0)}-6\bigr)}{36f_0k_0^2}}$ and $m_\phi=0$, implying that $\xi=\frac{af_0({\bf H}^{(0)})^2}{12\pi^2}$, $\eta=0$\,, $g(\xi,\eta)=\frac{af_0({\bf H}^{(0)})^2+12\pi^2}{6|af_0({\bf H}^{(0)})^2-12\pi^2|}+\frac{1}{6}$  and ${\tilde{k}_{R,\phi}}^2=1-\frac{af_0({\bf H}^{(0)})^2}{12\pi^2}\,,\,0$. The first relation  (\ref{bound}) becomes

\begin{eqnarray}
\nonumber
&& \frac{8}{3}(m_Y\,r^*+1)\,F(m_Y R_\oplus)\,e^{-m_Y\,r^*}\,\lesssim\,0.008~,
\end{eqnarray}
hence the constraint on $m_Y$ imposed from GBP is

\begin{eqnarray}
\nonumber
m_Y>7.1\times 10^{-5}{\rm m}^{-1}~,
\end{eqnarray}
whereas the LARES experiment (\ref{boundLT2}) implies

\begin{eqnarray}
\nonumber
m_Y>1.2\times 10^{-6} {\rm m}^{-1}~,
\end{eqnarray}
a bound similar to  the one obtained earlier on using binary pulsars~\cite{Nelson:2010ru}, or the Gravity Probe B data~\cite{LamSakSta}.
It is important to note that a much stronger limit, $m_Y >10^4 {\rm m}^{-1}$, has been  obtained using the torsion balance experiments~\cite{LamSakSta}.

In conclusion, using data form Gravity Probe B and LARES missions, we obtain similar constraints on $m_Y$; a result that one could have anticipated since both these experiments are designed to test the same type of physical phenomenon. However, by using the stronger constraint for $m_Y,$ namely $m_Y >10^4 {\rm m}^{-1}$, we observe that the modifications to the orbital parameters (\ref{OrbitalParameters}) induced by Noncommutative Spectral Geometry are indeed small, confirming the consistency between the predictions of NCSG as a gravitational theory beyond GR and the Gravity Probe B and LARES measurements. At this point let us stress that, in principle,  space-based experiments can be used to test parameters of fundamental theories.

\section{Conclusions}\label{conclusions}

In the context of Extended Gravity, we have studied the linearized field equations in the limit of weak gravitational fields and small velocities generated by rotating gravitational sources, aiming at constraining the free parameters, which can be seen as effectives masses (or lengths), using recent recent experimental results.  We have studied the precession of spin of a gyroscope orbiting about a rotating gravitational source. Such a gravitational field gives rise, according to GR predictions, to  geodesic and  Lense-Thirring processions, the latter being strictly related to the off-diagonal terms of the metric tensor generated by the rotation of the source. We have focused in particular on the gravitational field generated by the Earth, and on the recent experimental results obtained by the Gravity Probe B satellite, which  tested the geodesic and Lense-Thirring spin precessions with high precision. 

In particular, we have calculated the corrections of the precession induced by scalar, tensor and curvature  corrections. Considering an almost circular orbit,  we integrated the Gauss equations and obtained the variation of the parameters at first order with respect to the eccentricity. 
We have shown that the induced EG effects depend on the effective masses $m_R$, $m_Y$ and $m_\phi$ (\ref{par_sol}), while the nonvalidity of the Gauss theorem implies  that these effects also depend on the geometric form and size of the rotating source. Requiring that the corrections are within the experimental errors, we then imposed constraints on the free parameters of the considered EG model. Merging the experimental results of Gravity Probe B and LARES, our results can be summarized  as follows:

\begin{eqnarray}
&& g(\xi,\eta)\big(m_R\tilde{k}_R\,r^*+1)\,F(m_R\tilde{k}_R R_\oplus)\,e^{-m_R\tilde{k}_R\,r^*}+[1/3-g(\xi,\eta)](m_R\tilde{k}_\phi\,r^*+1)\,F(m_R\tilde{k}_\phi R_\oplus)\,e^{-m_R\tilde{k}_\phi\,r^*}
\nonumber\\\nonumber\\
&&\qquad\qquad\qquad\qquad\qquad\qquad\qquad\qquad\qquad+\frac{8}{3}(m_Y\,r^*+1)\,F(m_Y R_\oplus)\,e^{-m_Y\,r^*}\,\lesssim\, 0.008~,
\end{eqnarray}
and
\begin{eqnarray}
m_Y\geq 1.2\times 10^{-6} m^{-1}~.
\end{eqnarray}
It is interesting to note that the field equation for the potential $A_i$, Eq.~(\ref{PMfieldequationFOG3}c),
is time-independent provided the potential $\Phi$ is time-independent. This aspect guarantees that the solution Eq.~(\ref{sol_A_pointlike}) does not depend on the masses $m_R$ and $m_\phi$ and, in the case of $f(R,\,\phi)$ gravity, the solution is the same as in GR. 
In the case of spherical symmetry, the hypothesis of a radially static source is no longer considered, and the obtained solutions depend on choice of the $f(R,\,\phi)$ ET model, since  the geometric factor $F(x)$ is time-dependent. Hence in this case,  gravito-magnetic corrections to GR  emerge   with  time-dependent sources.

A final remark deserves the case of Noncommutative Spectral Geometry  that we discussed above. This model  descends from a fundamental theory and can be considered as a particular case of Extended Gravity. Its parameters, can be probed in the weak-field limit and at local scales, opening new perspectives worth to be further developed~\cite{stequest,GINGER}.

\end{document}